\documentclass[12pt]{article}
\makeatletter\if@twocolumn\PassOptionsToPackage{switch}{lineno}\else\fi\makeatother

\usepackage{amsfonts,amssymb,amsbsy,latexsym,amsmath,tabulary,graphicx,times,caption,fancyhdr}
\usepackage{url,multirow,morefloats,floatflt,cancel,tfrupee}
\makeatletter

\AtBeginDocument{\@ifpackageloaded{textcomp}{}{\usepackage{textcomp}}}
\makeatother
\usepackage{colortbl}
\usepackage{xcolor}
\usepackage{pifont}
\usepackage[nointegrals]{wasysym}
\usepackage{bm}

\makeatletter

\def\mcWidth#1{\csname TY@F#1\endcsname+\tabcolsep}

\def\cAlignHack{\rightskip\@flushglue\leftskip\@flushglue\parindent\z@\parfillskip\z@skip}
\def\rAlignHack{\rightskip\z@skip\leftskip\@flushglue \parindent\z@\parfillskip\z@skip}

\@ifundefined{etal}{}{}

\usepackage{ifxetex}
\ifxetex\else\if@twocolumn\@ifpackageloaded{stfloats}{}{\usepackage{dblfloatfix}}\fi\fi

\AtBeginDocument{
\expandafter\ifx\csname eqalign\endcsname\relax
\def\eqalign#1{\null\vcenter{\def\\{\cr}\openup\jot\m@th
  \ialign{\strut$\displaystyle{##}$\hfil&$\displaystyle{{}##}$\hfil
      \crcr#1\crcr}}\,}
\fi
}

\AtBeginDocument{%
  \@ifpackageloaded{endfloat}%
   {\renewcommand\efloat@iwrite[1]{\immediate\expandafter\protected@write\csname efloat@post#1\endcsname{}}}{\newif\ifefloat@tables}%
}%

\def\BreakURLText#1{\@tfor\brk@tempa:=#1\do{\brk@tempa\hskip0pt}}
\let\lt=<
\let\gt=>
\def\processVert{\ifmmode|\else\textbar\fi}

\@ifundefined{subparagraph}{
\def\subparagraph{\@startsection{paragraph}{5}{2\parindent}{0ex plus 0.1ex minus 0.1ex}%
{0ex}{\normalfont\small\itshape}}%
}{}

\newcommand\role[1]{\unskip}
\newcommand\aucollab[1]{\unskip}

\@ifundefined{tsGraphicsScaleX}{\gdef\tsGraphicsScaleX{1}}{}
\@ifundefined{tsGraphicsScaleY}{\gdef\tsGraphicsScaleY{.9}}{}
\def\checkGraphicsWidth{\ifdim\Gin@nat@width>\linewidth
	\tsGraphicsScaleX\linewidth\else\Gin@nat@width\fi}

\def\checkGraphicsHeight{\ifdim\Gin@nat@height>.9\textheight
	\tsGraphicsScaleY\textheight\else\Gin@nat@height\fi}

\def\fixFloatSize#1{}
\let\ts@includegraphics\includegraphics

\def\inlinegraphic[#1]#2{{\edef\@tempa{#1}\edef\baseline@shift{\ifx\@tempa\@empty0\else#1\fi}\edef\tempZ{\the\numexpr(\numexpr(\baseline@shift*\f@size/100))}\protect\raisebox{\tempZ pt}{\ts@includegraphics{#2}}}}

\AtBeginDocument{\def\includegraphics{\@ifnextchar[{\ts@includegraphics}{\ts@includegraphics[width=\checkGraphicsWidth,height=\checkGraphicsHeight,keepaspectratio]}}}

\DeclareMathAlphabet{\mathpzc}{OT1}{pzc}{m}{it}

\def\URL#1#2{\@ifundefined{href}{#2}{\href{#1}{#2}}}

\def\UrlOrds{\do\*\do\-\do\~\do\'\do\"\do\-}%
\g@addto@macro{\UrlBreaks}{\UrlOrds}

\edef\fntEncoding{\f@encoding}

\makeatother

\newif\ifmultipleabstract\multipleabstractfalse%
\makeatletter

\def\wileyIndent{1pt}
\usepackage[a4paper,margin=2cm,headsep=.5cm,top=2.5cm,right=2.5cm,left=2.5cm,bottom=2.5cm]{geometry}

\renewenvironment{abstract}
{\vspace*{-1pc}\trivlist\item[]\leftskip\wileyIndent\hrulefill\par\vskip4pt\noindent\textbf{\abstractname}\mbox{\null}\\}{\par\noindent\hrulefill\endtrivlist}

\def\author#1{\gdef\@author{\hskip-\dimexpr(\tabcolsep)\hskip\wileyIndent\parbox{\dimexpr\textwidth-\wileyIndent}{\raggedright#1}}}

\def\title#1{\gdef\@title{\raggedright\bfseries\ifx\@articleType\@empty\else\@articleType\\\fi#1}}

\let\@articleType\@empty \def\articletype#1{\gdef\@articleType{{\normalfont\itshape#1}}}

\fancypagestyle{headings}{\fancyhf{}\fancyhead[C]{\RunningHead}\fancyfoot[C]{\thepage}}

\date{}

\def\NormalBaseline{\def\baselinestretch{1.1}}
\usepackage[noindentafter]{titlesec}
\titleformat{\section}[hang]{\NormalBaseline\filright\bfseries\boldmath}
{\large\thesection.\hspace{-6pt}}
{10pt}
{}
[]
\titleformat{\subsection}[hang]{\NormalBaseline\filright\bfseries\boldmath}
{\thesubsection.\hspace{-6pt}}
{10pt}
{}
[]
\titleformat{\subsubsection}[hang]{\NormalBaseline\filright\itshape\boldmath}
{\thesubsubsection.\hspace{-6pt}}
{10pt}
{}
[]
\titleformat{\paragraph}[runin]{\NormalBaseline\filright\itshape\boldmath}
{\theparagraph\hspace{-6pt}}
{10pt}
{}
[]
\titleformat{\subparagraph}[runin]{\NormalBaseline\filright\boldmath}
{\thesubparagraph\hspace{-6pt}}
{10pt}
{}
[]

\titlespacing{\section}{0pt}{1.5\baselineskip}{.2\baselineskip}
\titlespacing{\subsection}{0pt}{1\baselineskip}{.2\baselineskip}
\titlespacing{\subsubsection}{0pt}{.5\baselineskip}{10pt}
\titlespacing{\paragraph}{0pt}{.5\baselineskip}{10pt}
\titlespacing{\subparagraph}{0pt}{.5\baselineskip}{10pt}

\makeatother

\usepackage[T1]{fontenc}
\makeatother
\usepackage[numbers,sort&compress,super]{natbib}

\bibpunct{[}{]}{,}{s}{}{}
\def\thanksspace{{\phantom{\textsuperscript{\thefootnote}}}}

\usepackage{float}

\begin{document}

\title{Enhancement of Ising superconductivity in monolayer NbSe$_2$ via surface fluorination}

\author{\textit{Jizheng Wu}\textsuperscript{1}, \textit{Wujun Shi}\textsuperscript{2}, \textit{Chong Wang}\textsuperscript{3}, \textit{Wenhui Duan}\textsuperscript{3,4}, \textit{Yong Xu}\textsuperscript{3,4}, and \textit{Chen Si}\textsuperscript{1}\thanks{E-mail:sichen@buaa.edu.cn}{\thanksspace}\space \\[-3pt]\normalsize\normalfont ~\\
\textsuperscript{1}{School of Materials Science and Engineering\unskip,  Beihang University\unskip, Beijing\unskip, 100191\unskip, China}~\\
\textsuperscript{2}{Center for Transformative Science, Shanghai High Repetition Rate XFEL and Extreme Light Facility (SHINE), ShanghaiTech University, Shanghai 201210, China}~\\
\textsuperscript{3}{State Key Laboratory of Low Dimensional Quantum Physics and Department of Physics, Tsinghua University, 100084, Beijing, China}~\\
\textsuperscript{4}{Frontier Science Center for Quantum Information, Beijing 100084, China}}

\def\RunningHead{}\def\RunningAuthor{Jizheng \MakeLowercase{\textit{et al.}} }

\maketitle

\begin{abstract}
Recently discovered Ising superconductors have garnered considerable interest due to their anomalously large in-plane upper critical fields ($B_{c2}$). However, the requisite strong spin-orbital coupling in the Ising pairing mechanism generally renders these superconductors heavy-element dominant with notably low superconducting transition temperatures ($T_c$). Here, based on the Migdal-Eliashberg theory and the mean-field Bogoliubov-de Gennes Hamiltonian, we demonstrate a significant enhancement of Ising superconductivity in monolayer NbSe$_2$ through surface fluorination, as evidenced by concomitant improvements in $T_c$ and $B_{c2}$. This enhancement arises from three predominant factors. Firstly, fluorine atoms symmetrically and stably adhere to both sides of the monolayer NbSe$_2$, thereby maintaining the out-of-plane mirror symmetry and locking carrier spins out-of-plane. Secondly, fluorination suppresses the charge density wave in monolayer NbSe$_2$ and induces a van Hove singularity in the vicinity of the Fermi level, leading to a marked increase in the number of carriers and, consequently, strengthening the electron-phonon coupling (EPC). Lastly, the appearance of fluorine-related, low-frequency phonon modes further augments the EPC. Our findings suggest a promising avenue to elevate $T_c$ in two-dimensional Ising superconductors without compromising their Ising pairing.

\end{abstract}

\def\keywordstitle{Keywords}


\section{Introduction}

Highly crystalline two-dimensional (2D) superconductors present unparalleled opportunities for delving into novel quantum phenomena such as quantum metallic states\cite{saito2015metallic,tsen2016nature}, quantum Griffith singularities\cite{Quantum_Griffiths}, and superconducting field-effect transistors\cite{science1999,nature2008}. A recent notable advancement is the discovery of ``Ising pairing'' in 2D crystalline superconductors\cite{Nphys_NbSe2,NC_TaS2_NbSe2,Science_MoS2,saito2016superconductivity,PNAS_WS2,Nature_Materials_NbSe2,PdTe2_Nano_letters,Science_Sn,WangChong_PRL,PhysRevB.93.180501}. When a magnetic field is applied parallel to the plane of sufficiently thin 2D superconductors, the interaction between the magnetic field and electron orbitals is suppressed\cite{Nphys_NbSe2}. In this limit, the principal mechanism disrupting superconductivity is the Pauli paramagnetism, which sets the Pauli paramagnetic limit ($B_p$) to be the in-plane upper critical field ($B_{c2}$)\cite{Pauli_1,Pauli_2}. However, in recently discovered atomically-thin Ising superconductors---including intrinsic monolayer NbSe$_2$ \cite{Nphys_NbSe2}and TaS$_2$\cite{NC_TaS2_NbSe2}, monolayer MoS$_2$\cite{Science_MoS2,saito2016superconductivity}and WS$_2$ \cite{PNAS_WS2}facilitated with charge carriers via ionic liquid gating, alongside few-layer PdTe$_2$ \cite{PdTe2_Nano_letters}and stanene\cite{Science_Sn,WangChong_PRL}---the superconductivity persists even when exposed to an anomalously large in-plane magnetic field surpassing $B_p$.

In 2D Ising superconductors, electron spins within Cooper pairs are robustly anchored by an effective out-of-plane Zeeman magnetic field, engendered by a substantial spin-orbit coupling (SOC) in conjunction with either breaking of inversion symmetry (type I)\cite{NC_TaS2_NbSe2,Science_MoS2,saito2016superconductivity,PNAS_WS2,Nphys_NbSe2,Nature_Materials_NbSe2} or the presence of multiple degenerate orbitals (type II)\cite{PdTe2_Nano_letters,Science_Sn,WangChong_PRL}. Consequently, the Cooper pairs are insensitive to the in-plane external magnetic field. However, given that large SOC is indispensable for the formation of Ising pairing, the Ising superconductors are generally composed of heavy elements which tend to show weak electron-phonon coupling (EPC). As a result, the identified Ising superconductors all have notably low critical transition temperatures ($T_c$), especially the intrinsic ones. For instance, monolayer NbSe$_2$ and TaS$_2$ have $T_c$ values of merely about 3 K\cite{NC_TaS2_NbSe2}, while few-layer PdTe$_2$ and stanene possess even lower $T_c$ values of approximately 700 mK\cite{PdTe2_Nano_letters} and 1 K\cite{Science_Sn}, respectively. To increase the $T_c$, integrating a substantial amount of light elements within the materials is a viable approach\cite{H_MgB2_PRL}. However, such an approach often causes structural instability. For instance, high-$T_c$, hydrogen-rich materials can only stabilize under extreme pressure\cite{H3S,PH_JACS,MaYanming_2022}. Additionally, the introduction of lighter atoms is highly likely to alter the  original lattice symmetry, thereby destroying the Ising configuration of carrier spins. As a result, elevating the $T_c$ of Ising superconductors without compromising the fragile Ising pairing continues to pose a significant challenge.

In this study, we demonstrate that surface fluorination serves as an effective strategy for substantially enhancing Ising superconductivity in monolayer NbSe$_2$. Upon fluorination, fluorine atoms adsorb symmetrically and stably onto both sides of the NbSe$_2$ monolayer, thereby preserving its mirror symmetry and locking carrier spins in the out-of-plane direction. Importantly, surface fluorination enhances the EPC in monolayer NbSe$_2$ through two primary mechanisms. Initially, the fluorine adatoms suppress the charge density wave (CDW) and induce a van Hove singularity (VHS) within the density of states (DOS) proximate to the Fermi level ($E_F$), collectively increasing the number of carriers and thereby amplifying the EPC. Subsequently, the advent of low-frequency, strong-coupling phonon modes associated with fluorine further intensifies the EPC. By solving anisotropic Migdal-Eliashberg equations\cite{Migdal,Eliashberg} and the Bogoliubov-de Gennes (BdG) equation from first principles, we find that fluorination not only triples the $T_c$ but also markedly increases the in-plane $B_{c2}$ of monolayer NbSe$_2$.

\section{Method Section}

Our calculations were performed within the density functional theory (DFT) and density functional perturbation theory (DFPT)\cite{DFPT}, utilizing the Quantum Espresso package\cite{QE}. We employed norm-conserving pseudopotentials\cite{ONCV} and a plane-wave energy cutoff of 100 Ry. The exchange-correlation interactions were characterized by generalized gradient approximation of Perdew-Burke-Ernzerhof functional\cite{PBE}. For the monolayer model, a vacuum layer with a thickness greater than 15 \AA{} was introduced to avoid interaction between adjacent images. All structures underwent full optimization until the forces exerted on each ion fell below $10^{-5}$ Ry/Bohr. The electronic and phononic Brillouin zones for functionalized NbSe$_2$ in the $1 \times 1$ phase (pristine NbSe$_2$ in the $3 \times 3$ NbSe$_2$ CDW phase) were sampled using $18 \times 18$ ($6 \times 6$) $\textbf{k}$-grids and $9\times 9$ ($3 \times 3$) $\textbf{q}$-grids, respectively. Spin-orbit coupling effects were accounted for into our computations for both electronic structures and lattice dynamics. The superconducting gap and critical temperature were determined through the anisotropic Migdal-Eliashberg approach\cite{EPW_ME} as operationalized within the EPW code\cite{EPW_1,EPW_2}. Electron-phonon matrix elements were interpolated on 240 $\times$ 240 (80 $\times$ 80) $\textbf{k}$-point grids and 120 $\times$ 120 (40$\times$40) $\textbf{q}$-point grids for NbSe$_2$F$_2$ (the 3$\times$3 NbSe$_2$ CDW phase). We employed a Matsubara frequency cutoff of 0.4 eV. The Dirac deltas were substituted by Lorentzian functions with widths of 12.5 meV for electrons and 0.50 meV for phonons. The effective Coulomb interactions between Cooper pairs were described using a Morel-Anderson pseudopotential $\mu^*$, set to 0.15.

We determined the spin susceptibility $\chi$ using the collinear fixed-spin moment (FSM) calculations, alternatively known as the constrained local moments approach\cite{PRX_Mazin,NC_Mazin}. 
In these calculations, we constrained the size of the magnetic moment on the Nb atom. This approach enabled us to determine of the total energy relative to the nonmagnetic reference state as a function
of the total magnetization $m$. Subsequently, we fitted the results using the expression $E(m)=a_0+a_1 m^2+a_2 m^4+a_3 m^6+a_4 m^8+\cdots$, where $E(m)$ denotes the total energy for a given magnetization $m$. This fitting enabled us to compute $\chi = 1/a_1$. To obtain converged values of $\chi$, we employed a $28\times28$ ($6\times6$) $\textbf{k}$-grid for the bare and functionalized NbSe$_2$ in the $1 \times 1$ phases (the $3 \times 3$ CDW phase of NbSe$_2$). We also set an energy convergence threshold of $10^{-9}$ Ry and utilized as many as 61 points to plot energy against magnetization within the range of 0 to 0.6 $\mu$B.

\section{Results and Discussion}
We begin our study with the CDW phase of monolayer NbSe$_2$. As the temperature decreases, monolayer NbSe$_2$ experiences a second-order phase transition to a $3 \times 3$ CDW state, subsequently transitioning to a superconducting state that coexists with the CDW order\cite{NbSe2_CDW_STM,Nature_nano_NbSe2_CDW}. Hence in essence the Ising superconductivity of monolayer NbSe$_2$ emerges in its CDW phase. However, previous theoretical studies considering Ising superconductivity of monolayer NbSe$_2$ were all based on its high-symmetry $1\times1$ phase\cite{PhysRevB.93.180501,Nphys_NbSe2,NC_Mazin,NC_TaS2_NbSe2}. In the following discussion, we will show that the CDW phase not only supports the Ising pairing but also possesses a $T_c$ value that is very close to the experimental measurement for monolayer NbSe$_2$. In contrast, calculations on the high-symmetry phase of monolayer NbSe$_2$ yielded a $T_c$ that deviates significantly from the experimental value, as demonstrated in a previous work\cite{JiFeng_PRB}.

Figure 1a depicts the $3\times3$ CDW structure for monolayer NbSe$_2$. This structure has been identified as the ground-state CDW phase, as corroborated by the calculated scanning tunneling microscopy images and Raman CDW modes\cite{CS_Nanoletters,FJ_PRB_NbSe2_CDW}, which are consistent with experimental results\cite{NbSe2_CDW_STM,Nature_nano_NbSe2_CDW}. In contrast to the high-symmetry phase possessing uniform Nb-Nb distances, the CDW phase features distortions that lead to the emergence of triangular Nb trimers and hexagonal Nb clusters. Importantly, these CDW-induced distortions do not violate the out-of-plane mirror symmetry of monolayer NbSe$_2$, as they primarily occur in the Nb atomic plane and the displacements in the Se atoms are negligible. This preserved mirror symmetry confines the crystal field $\textbf{E}$ to the basal plane. As such, for the electron motion with momentum $\textbf{k}$ within this plane, the antisymmetric SOC arising from the breaking of inversion symmetry engenders an effective out-of-plane magnetic field $\textbf{B}_{\text{so}} \propto \textbf{E} \times \textbf{k}$. Thus, akin to the high-symmetry phase, the electron spins in the CDW phase remain locked in the out-of-plane direction, facilitating Ising Cooper pairing between a spin-up electron with momentum $\textbf{k}$ at the Fermi surface and its time-reversed counterpart with opposite spin and momentum.

\bgroup
\fixFloatSize{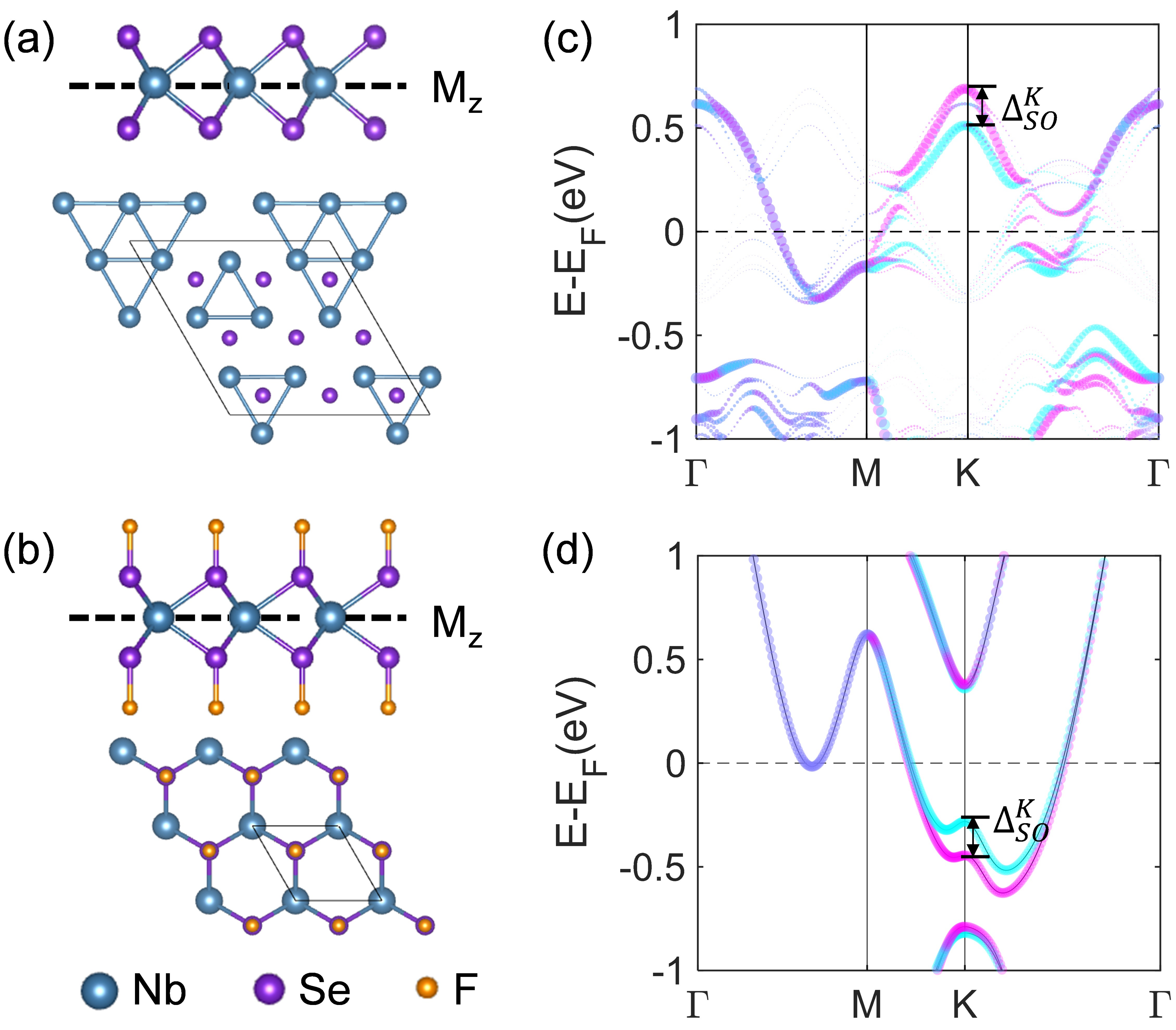}
\begin{figure*}[t!]
\centering \makeatletter\IfFileExists{Figure1.pdf}{\includegraphics[width=1.0\linewidth]{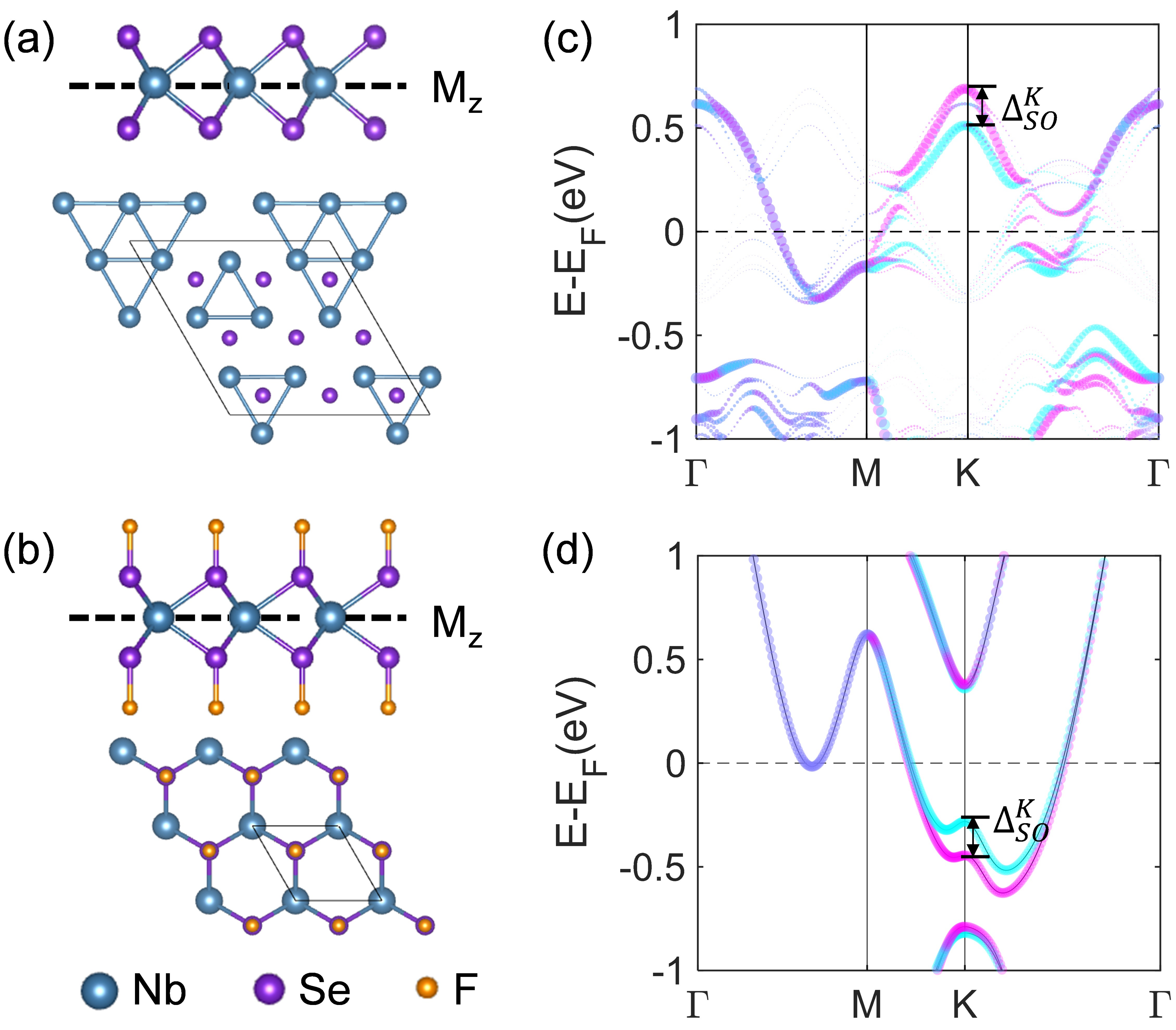}}{}
\makeatother
\caption{Lattice structures and band structures for the monolayer NbSe$_2$ in the $3\times3$ CDW phase (a, c) and the monolayer NbSe$_2$F$_2$ (b, d). In (a) and (b), side and top views of lattice structures are both shown. In (c), energy bands of the $3\times3$ CDW phase of monolayer NbSe$_2$ are unfolded into the $1\times1$ Brillion zone. The blue and purple lines in (c) and (d) represent the energy bands with pure $m_z = 1$ character (spin-up) and pure $m_z = -1$ character (spin-down), respectively.}
\label{Figure1}
\end{figure*}
\egroup

To enhance the $T_c$ of monolayer NbSe$_2$ without sacrificing the Ising pairing, maintaining the out-of-plane mirror symmetry is a good choice. We therefore propose surface functionalization of monolayer NbSe$_2$ in its CDW phase using lightweight fluorine atoms, based on two primary considerations. First, monolayer NbSe$_2$ with top and bottom surfaces fully fluorinated (Figure 1b) is thermodynamically more favorable compared to configurations with partial functionalization (see Figure S1 in the Supporting Information). Moreover, fluorine atoms are symmetrically adsorbed on both sides of monolayer NbSe$_2$, specifically aligning atop the selenium atoms, which effectively conserves the material's mirror symmetry. Second, the fully fluorinated monolayer NbSe$_2$ is dynamically stable, as verified by the absence of imaginary-frequency phonons (see Figure 3a). In contrast, a fully hydrogenated monolayer NbSe$_2$ displays marked structural instability, as indicated by the presence of numerous strongly soft modes throughout almost the entire Brillouin zone (Figure S2 in the Supporting Information).

Henceforth, fully fluorinated monolayer NbSe$_2$, characterized by the stoichiometry of NbSe$_2$F$_2$, will be referred to as NbSe$_2$F$_2$. Notably, compared with pristine monolayer NbSe$_2$ with a $3\times3$ CDW order, NbSe$_2$F$_2$ exhibits a suppressed CDW order, with the lattice reduced to a \(1 \times 1\) periodicity (see Figure 1b and Figure S3 in the supported information). Changes of CDW order by adsorption have also been observed in 1T-TaS$_2$ where water adsorption induces a $\sqrt{13} \times \sqrt{13}$ to $3\times3$ CDW order transition\cite{TaS2_Nanoletters}. In NbSe$_{2}$F$_{2}$, the bond lengths are as follows: Se-F at 1.84 \AA, and Nb-Se at 2.59 \AA, which is comparable to the 2.59-2.62 \AA\ Nb-Se bond lengths observed in the CDW phase of monolayer NbSe$_{2}$. However, the Nb-Nb distance expands from 3.34-3.42 \AA\ in the CDW phase of monolayer NbSe$_{2}$ to 3.67 \AA\ in NbSe$_{2}$F$_{2}$, attributable to lattice expansion induced by fluorine adsorption (see the structural details of NbSe$_2$F$_2$ and the CDW phase of monolayer NbSe$_{2}$ in Table S1 and S2 of the Supporting Information). Given the prior successful synthesis of fully fluorinated graphene through treatment with XeF$_2$ or F$_2$ gas\cite{AS_2016}, it is reasonable to anticipate that NbSe$_2$F$_2$ could be produced through a similar synthetic methodology.

Figures 1c and 1d illustrate the electronic structures of pristine and fluorinated monolayer NbSe$_2$.  To facilitate a direct comparison, the energy bands of pristine NbSe$_2$ in the \(3 \times 3\) CDW phase are unfolded into the \(1 \times 1\) Brillouin zone. Both systems display pronounced spin splitting at the K point ($\Delta_{SO}^K$) near the Fermi level, induced by SOC. The $\Delta_{SO}^K$ values in NbSe$_2$ and NbSe$_2$F$_2$ are closely aligned, measuring 0.177 and 0.164 eV, respectively. These values correspond to out-of-plane Zeeman fields of \(3.06 \times 10^3\) and \(2.83 \times 10^3\) Tesla, respectively. 

Significantly, NbSe$_2$F$_2$ features a larger DOS at the Fermi level ($N_F$), compared to monolayer NbSe$_2$ in the CDW phase--registering 2.49 states/eV/NbSe$_2$F$_2$ versus 1.67 states/eV/NbSe$_2$, a condition favorable for superconductivity (see Figures 2a and 2b). This $N_F$ enhancement primarily arises from the suppression of CDW in NbSe$_2$F$_2$ and the concomitant formation of a VHS in the electronic DOS around $E_F$. In pristine NbSe$_2$, the emergence of CDW leads to partial energy gap opening at some k-points on the Fermi surface. 
As depicted in Figure 2c, the Fermi surfaces of monolayer NbSe$_2$ in the high-symmetry 1 × 1 1H phase and in the 3 × 3 CDW phase are compared, with the Fermi surface of the CDW phase unfolded into the 1 × 1 Brillouin zone. This comparison clearly shows partial gapping of the K- and K’-centered Fermi contours post-CDW formation. Consequently, in the DOS for monolayer NbSe$_2$ in the CDW phase, $E_F$ is situated within a quasi-energy gap (Figure 2a), resulting in a small $N_F$. However, in NbSe$_2$F$_2$, not only is the CDW order suppressed, but a VHS appears also near $E_F$ (Figure 2b), contributing to a significantly increased $N_F$. Figure 2d presents a three dimensional plot of the energy band crossing the $E_F$ for NbSe$_2$F$_2$, where a saddle point close to $E_F$, labeled as the Q point, is visible. According to the symmetry of the Brillouin zone, there are six saddle points in the first Brillouin zone, all sharing the same energy as Q; however, for clarity, only Q is marked in Figure 2d. It is these saddle points that lead to the formation of VHS close to $E_F$. Similar phenomenon has observed in many 2D materials\cite{JPCL_2018}. 

\bgroup
\fixFloatSize{Figure2.pdf}
\begin{figure*}[t!]
\centering \makeatletter\IfFileExists{Figure2.pdf}{\includegraphics[width=1.0\linewidth]{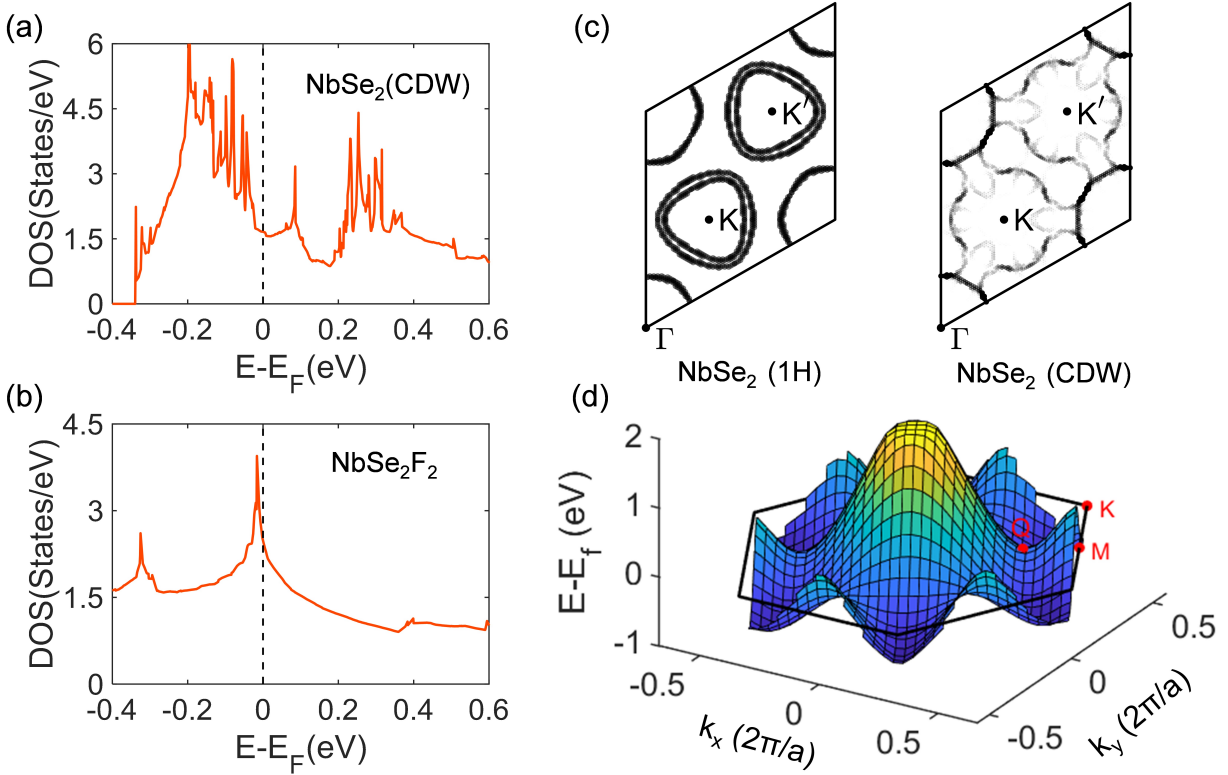}}{}
\makeatother
\caption{{Density of states for the monolayer NbSe$_2$ in the $3\times3$ CDW phase (a) and in the monolayer NbSe$_2$F$_2$ (b). (c) Fermi surfaces of monolayer NbSe$_2$ in the high-symmetry 1H phase and in the CDW phase. (d) Three dimensional plot of the energy band crossing the $E_F$ for NbSe$_2$F$_2$. Q is a saddle point close to $E_F$.}}
\label{Figure2}
\end{figure*}
\egroup

The increase in $N_F$ of NbSe$_2$F$_2$ implicates an enhancement in EPC. Therefore, we further perform DFPT\cite{DFPT} calculations to assess the vibrational characteristics and EPC in both the $3 \times 3$ CDW phase of NbSe$_2$ and the $1 \times 1$ phase of NbSe$_2$F$_2$. Given the substantial impacts of SOC on the electronic structures of these systems, SOC is also included into the lattice dynamics calculations. Figures 3b and 3c depict the isotropically averaged Eliashberg spectral function $\alpha^2 F(\omega)$ and cumulative EPC $\lambda_{\text{iso}}(\omega)$ for each system respectively. The function \(\alpha^2 F(\omega)\) quantifies the mean interaction between phonons of energy \(\omega\) and electrons on the Fermi surface. This function is defined by the equation \(\alpha^2 F(\omega) = \frac{1}{N_F N_k N_q} \sum_{m,n} \sum_{q,\upsilon} \delta(\omega - \omega_{q\upsilon}) \times \sum_k |g_{k+q,k}^{q\upsilon,mn}|^2 \delta(E_{k+q,m} - E_F) \delta(E_{k,n} - E_F)\). In this context, \(\omega\) represents the phonon energy indexed by wave vector \(q\) and mode number \(\upsilon\); \(E\) denotes the electron energy, indexed by momentum \(k\) and band index \(m\) or \(n\); \(g_{k+q,k}^{q\upsilon,mn}\) indicates the electron-phonon coupling matrix element; \(N_F\) refers to the density of states at \(E_F\); and \(N_k\) and \(N_q\) represent the counts of \(k\) and \(q\) points, respectively. The cumulative EPC, \(\lambda_{\text{iso}}(\omega)\), is calculated using the formula \(\lambda_{\text{iso}}(\omega) = 2 \int_0^\omega \frac{\alpha^2 F(\omega')}{\omega'} d\omega'\). Notably, the total isotropic EPC constant reaches a high value of $\lambda_{\text{iso}} = 1.18$ in Nb$_2$Se$_2$F$_2$, as opposed to $\lambda_{\text{iso}} = 0.78$ in pristine NbSe$_2$ CDW phase. It is clearly seen that the elevated $\lambda_{\text{iso}}$ in Nb$_2$Se$_2$F$_2$ predominantly arises from the contribution of low frequency phonons in the energy region below 15 meV, accounting for approximately 80\% of the total $\lambda_{\text{iso}}$ (Figure 3c). Examination of the phonon DOS shown in Figure 3d further reveals significant contributions of fluorine vibrations in this energy range, which demonstrates strong hybridization with Nb and Se atomic vibrations. Given the $1/\omega$ factor in $\lambda_{\text{iso}}(\omega)$, the low-frequency fluorine phonons will significantly strengthen EPC. Hence, the substantial increase in $\lambda$ induced by surface fluorination arises from a synergistic effect of an increased DOS near $E_F$ and the emergence of fluorine-related strong-coupling phonon modes in the low-frequency regime.
\bgroup
\fixFloatSize{Figure3.pdf}
\begin{figure*}[t!]
\centering \makeatletter\IfFileExists{Figure3.pdf}{\includegraphics[width=\linewidth]{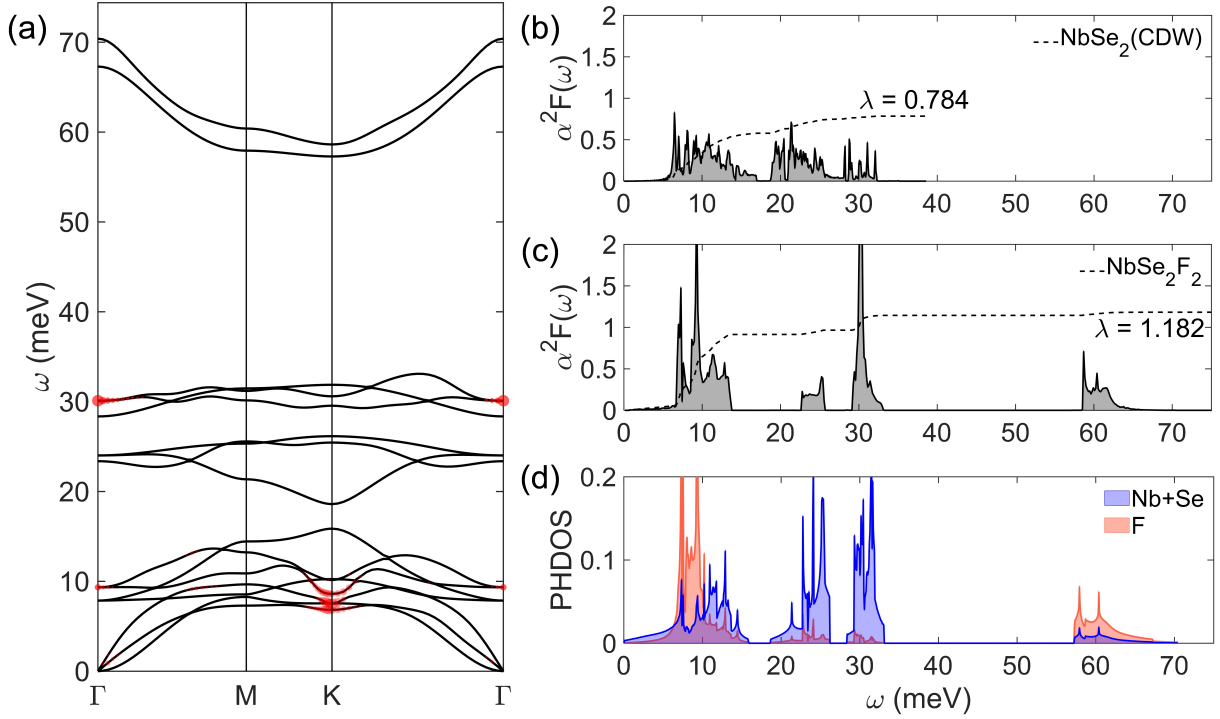}}{}
\makeatother
\caption{{(a) Phonon dispersion of NbSe$_2$F$_2$ where the size of red dots indicates the EPC strength of a given phonon mode ($\lambda_q^\upsilon$). (b) Isotropic Eliashberg function $\alpha^2 F(\omega)$ (solid black curve) and the cumulative EPC $\lambda_{\text{iso}} (\omega) = 2 \int_0^\omega \frac{\alpha^2 F(\omega')}{\omega'} d\omega'$ (dashed black curve) for the CDW phase of monolayer NbSe$_2$. (c) Isotropic Eliashberg function for NbSe$_2$F$_2$. (d) Atom-resolved phonon density of states for NbSe$_2$F$_2$.}}
\label{Figure3}
\end{figure*}
\egroup

Prior to evaluating the superconducting properties of both pristine and fluorinated NbSe$_2$, it is imperative to scrutinize the role of spin fluctuations in these systems, as such fluctuations have been demonstrated to serve as a potential source for Cooper pair-breaking in many materials\cite{PRX_Mazin,NPJ_2023}. To quantify the strength of spin fluctuations, we calculated the spin susceptibility $\chi$, employing the collinear FSM method\cite{PRX_Mazin,NC_Mazin}. In the FSM calculations, we impose a constraint on the magnetization and determine the total energy, $E$, under different total magnetic moments, $m$. From these calculations, we can establish $\chi$ to be $\chi = \left( \frac{\partial^2 E}{\partial m^2} \right)^{-1}$. Figure 4a presents the FSM calculations for monolayer NbSe$_2$ in both the high-symmetry and CDW phases, as well as for monolayer NbSe$_2$F$_2$. In each system, the total energy ascends monotonically with an increasing magnetic moment. Fitting the computed total energy, $E$, to the polynomial expression $E = a_0 + a_1 m^2 + a_2 m^4 + a_3 m^6 + a_4 m^8 + \cdots$, yields $\chi = 1/a_1$. Previous investigations have proposed that monolayer NbSe$_2$ in its high-symmetry phase exhibits pronounced spin fluctuations and is in proximity to a ferromagnetic instability\cite{PRX_Mazin}. It is noteworthy that our calculations reveal a substantial attenuation of spin fluctuations in both the CDW phase of monolayer NbSe$_2$ and in NbSe$_2$F$_2$. As illustrated in Figure 4b, the spin susceptibilities for these latter two systems are approximately 80\% and 90\% less than that of high-symmetry phase of monolayer NbSe$_2$, respectively.
\bgroup
\fixFloatSize{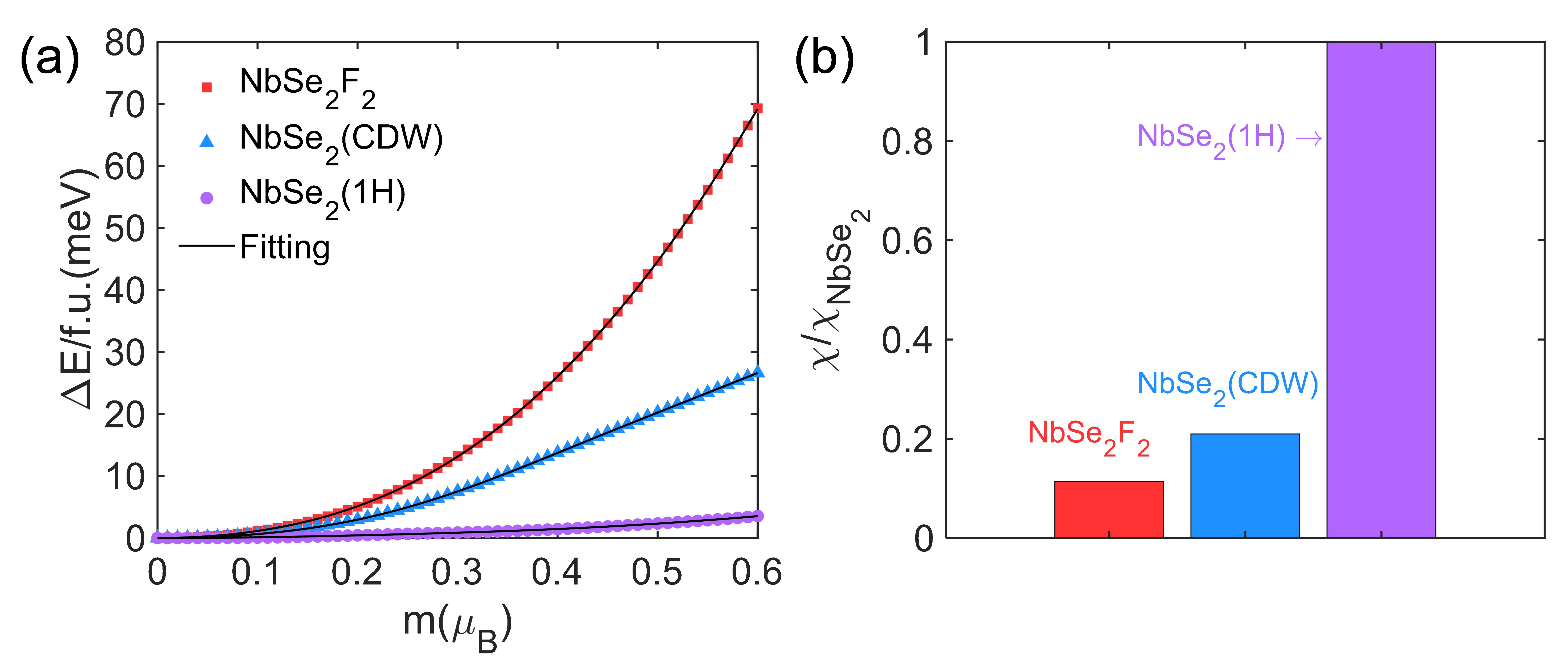}
\begin{figure*}[h!]
\centering \makeatletter\IfFileExists{Figure4.pdf}{\includegraphics[width=\linewidth]{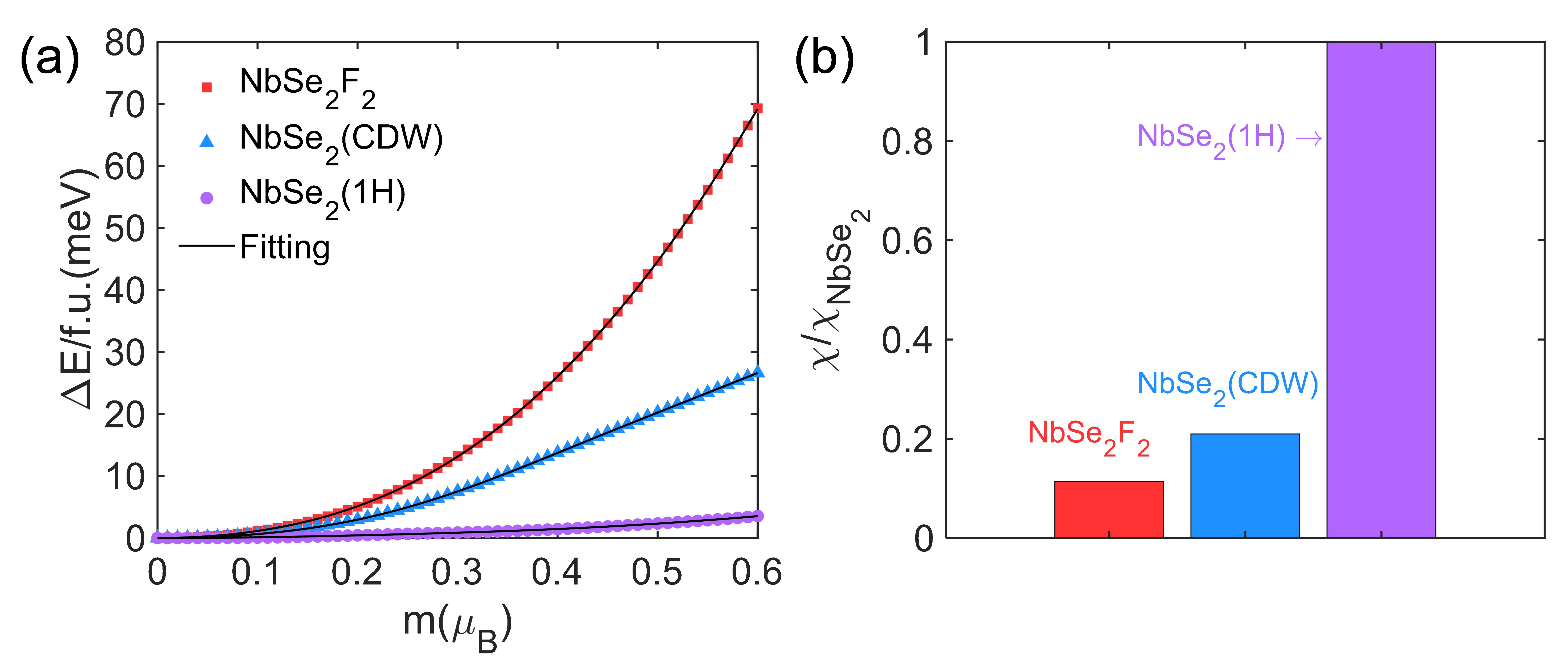}}{}
\makeatother
\caption{{(a) Change in total energy per formula unit as a function of magnetic moment for the high-symmetry $1H$ phase (purple circles), the CDW phase (blue triangles) of monolayer NbSe$_2$, and monolayer NbSe$_2$F$_2$ (red squares), calculated using collinear FSM methods relative to their respective nonmagnetic states. The black solid lines are polynomial fits of the calculated data. (b) Spin susceptibilities $\chi$ for the high-symmetry $1H$ phase of NbSe$_2$ (purple rectangles), the CDW phase of NbSe$_2$ (blue rectangles), and NbSe$_2$F$_2$ (red rectangles), each normalized by the $\chi$ value for the high-symmetry $1H$ phase of NbSe$_2$.}}
\label{Figure4}
\end{figure*}
\egroup

Given the weak spin fluctuations in the monolayer NbSe$_2$ CDW phase and NbSe$_2$F$_2$, we can justifiably neglect the effects of spin fluctuations on the superconductivity and consider the Cooper paring to be entirely due to EPC. Consequently, we proceed to calculate the superconductivity properties through solving the anisotropic Migdal-Eliashberg equations, utilizing our $ab$ $initio$ results of electrons, phonons, and electron-phonon matrices, as input parameters. Figure 5a illustrates the calculated energy distribution of superconducting gaps on the Fermi surface of NbSe$_2$F$_2$ across varying temperatures, juxtaposed against those of the NbSe$_2$ CDW phase. The resulting critical superconducting temperature $T_c$ for the NbSe$_2$ CDW monolayer is 3.8 K, which closely agrees with the experimental value of 3 K\cite{NC_TaS2_NbSe2,Nphys_NbSe2}. In stark contrast, NbSe$_2$F$_2$ exhibits a significantly elevated $T_c$ of 11.5 K, approximately threefold greater than the 3.8 K calculated for the NbSe$_2$ CDW monolayer.

\bgroup
\fixFloatSize{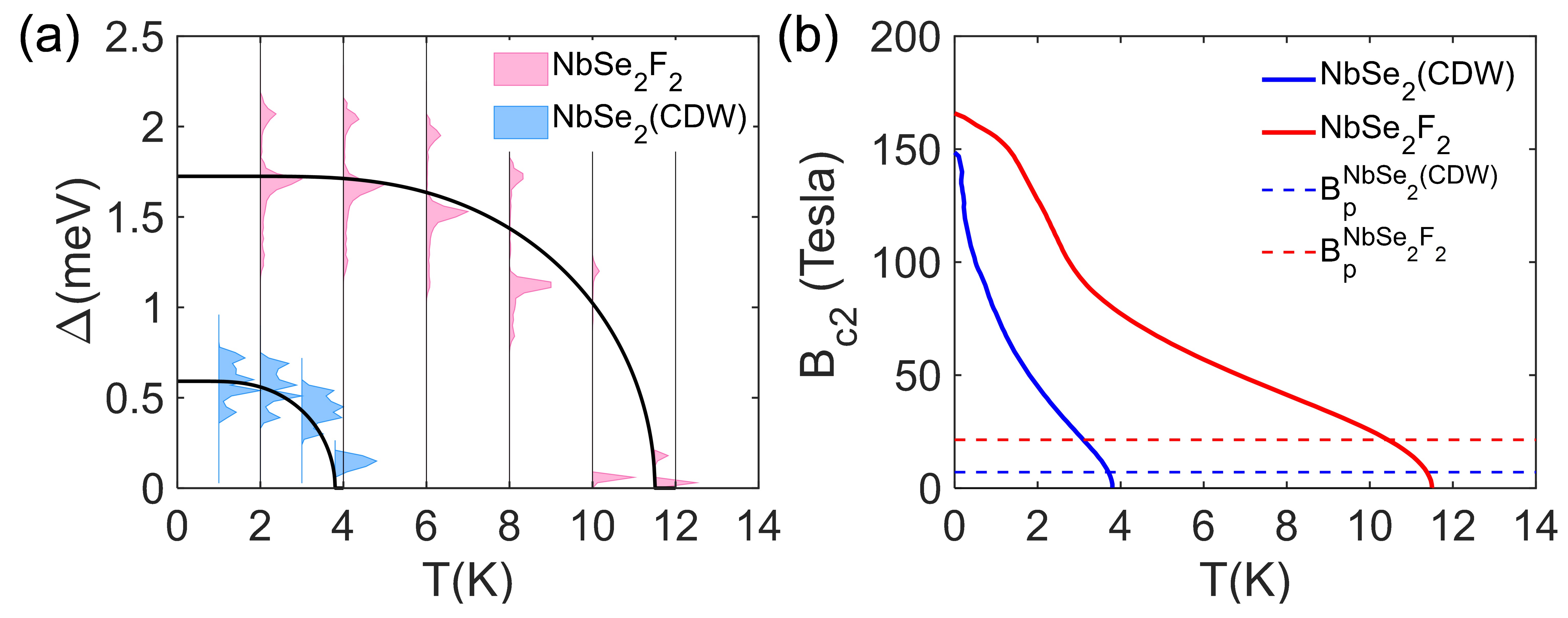}
\begin{figure*}[h!]
\centering \makeatletter\IfFileExists{Figure5.pdf}{\includegraphics[width=\linewidth]{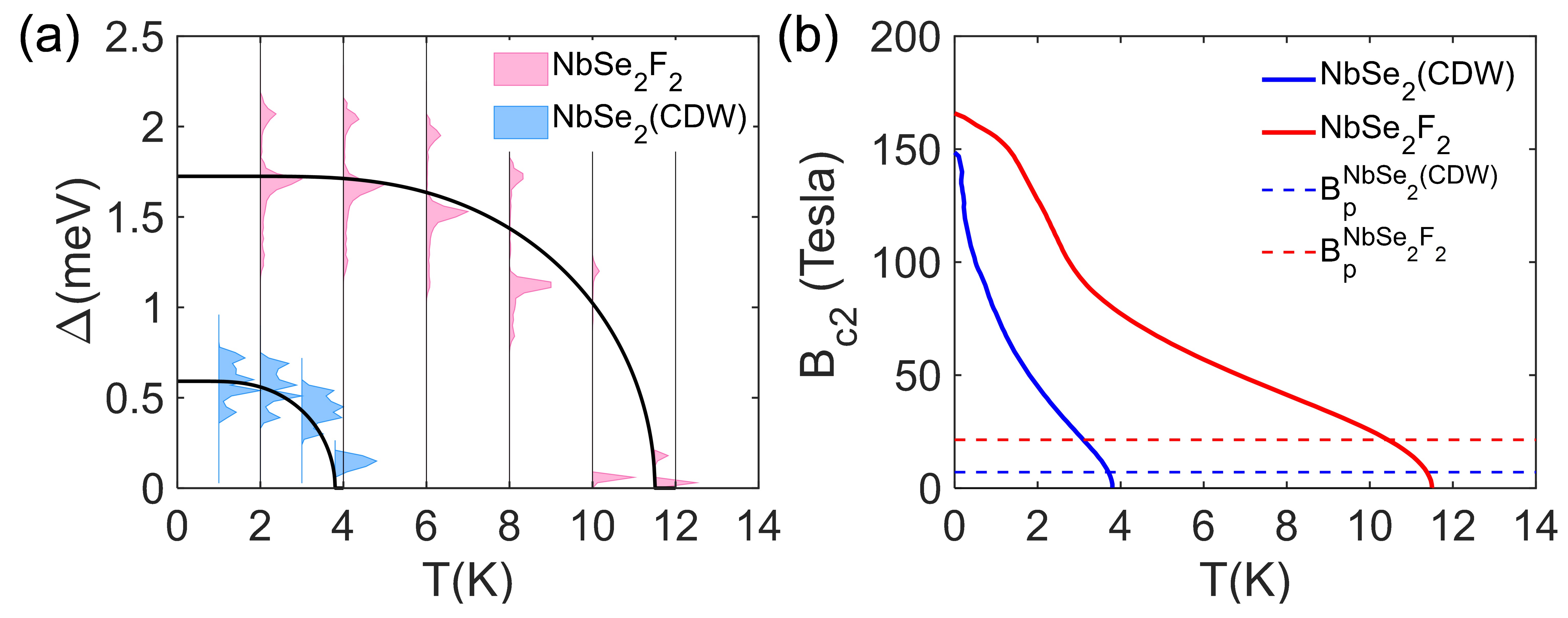}}{}
\makeatother
\caption{{(a) Energy distributions of the superconducting gaps $\Delta_k$ of NbSe$_2$F$_2$ (red shaded area) and the CDW phase of NbSe$_2$ (blue shaded area). The solid lines represent the BCS-model fittings of the calculated data. (b) The temperature-dependent in-plane critical magnetic field $B_{c2}$ for NbSe$_2$F$_2$ (red solid lines) and the CDW phase of NbSe$_2$ (blue solid lines). The red and blue dashed lines correspond to the Pauli limits for NbSe$_2$F$_2$ and the CDW phase of NbSe$_2$, respectively.}}
\label{Figure5}
\end{figure*}
\egroup

It is found that the temperature dependencies of the superconducting gaps for monolayer NbSe$_2$ in CDW phase and for NbSe$_2$F$_2$ can be well captured using the Bardeen--Cooper--Schrieffer (BCS) model. Within the BCS model, the variation of the superconducting gap \(\Delta\) with temperature is ascertained by self-consistently solving the gap equation\cite{BCS_PR} using the values of $\Delta_0$ and $T_c$ derived from our $ab$ $initio$ calculations:
\begin{equation}
\int_0^{(k_B \Theta_D)/\Delta_0} \tanh\left(\frac{\alpha_{\text{BCS}} \tilde{E}}{2t}\right) \frac{d\tilde{\varepsilon}}{\tilde{E}} = \ln\left[\frac{2k_B \Theta_D}{\Delta_0}\right] \label{eq:gap_function}
\end{equation}

Here, \( t \) represents the normalized temperature \( T/T_c \), \( \tilde{\varepsilon} \) is the normalized energy \( \varepsilon/\Delta_0 \), \( \tilde{\Delta} \) denotes the normalized gap \( \Delta/\Delta_0 \), and \( \tilde{E} \) corresponds to the reduced quasiparticle energy, defined as \( \sqrt{\tilde{\varepsilon}^2 + \tilde{\Delta}^2} \). The term \(k_B \Theta_D\) signifies the Debye phonon energy, which is set to \(1000\Delta_0\). The BCS fits are depicted as black solid lines in Figure 5a.  The derived ratios of $\Delta_0/k_B T_c$ are 1.805 for monolayer NbSe$_2$ in the CDW phase and 1.741 for NbSe$_2$F$_2$, aligning closely with the ideal BCS value of 1.764. Consequently, we can estimate the in-plane $B_{c2}$ by mean-field calculations based on the premise of s-wave potential for pairing, as elucidated below.

To simulate the temperature-dependent $B_{c2}$, we construct the BdG Hamiltonian for the superconducting state in the presence of an in-plane magnetic field applied parallel to the $x$-axis by using the electronic Hamiltonian $H_N(\textbf{k}, B)$ of the normal state:
\begin{equation}
H_{\text{BdG}}(\textbf{k}) =
\begin{pmatrix}
H_N(\textbf{k}, B) & i\Delta \sigma_y \\
-\Delta i \sigma_y & -H_N^*(\textbf{k}, B)
\end{pmatrix}
\end{equation}
\begin{sloppypar}
Here, $H_N(\mathbf{k}, B) = H_N(\mathbf{k}) + H_z$. To derive $H_N(\mathbf{k})$, the real-space Hamiltonian $H_N(\mathbf{R})$ is initially obtained by fitting the \textit{ab initio} band structure incorporating SOC using the Wannier 90 code \cite{Wannier_1, Wannier_2}. This process employ the basis of Wannier Functions $\Psi = \left( \phi_{1\uparrow}, \phi_{2\uparrow}, \ldots, \phi_{i\uparrow}, \ldots, \phi_{M\uparrow}, \phi_{1\downarrow}, \phi_{2\downarrow}, \ldots, \phi_{i\downarrow}, \ldots, \phi_{M\downarrow} \right)$ , where i represents the orbital index and the total number of Wannier functions is 2M. Following this, $H_N(\mathbf{k})$ is computed via the Fourier transformation of $H_N(\mathbf{R})$, using $h_{ij}(\mathbf{k}) = \sum_{\mathbf{R}} h_{ij}^{R} \times e^{i\mathbf{k} \cdot \mathbf{R}}$, where $h_{ij}(\mathbf{k})$ and $h_{ij}^{\mathbf{R}}$ are the matrix elements of $H_N(\mathbf{k})$ and $H_N(\mathbf{R})$, respectively. The Zeeman term $H_z = \frac{1}{2} g \mu_B B \sigma_x \otimes I_{M \times M}$ accounts for the paramagnetic effect with Landé g-factor $g = 2$. $\Delta$ is the superconducting gap.
According to the $H_{\text{BdG}}(\textbf{k})$ in Eq. (2), the free energy density of the superconducting state is given by \cite{Nature_Materials_NbSe2,SOPC}:
\end{sloppypar}
\begin{equation}
F_s = \frac{|\Delta|^2}{U_0} - \frac{1}{\beta} \sum_{k,n} \ln \left(1 + e^{-\beta \epsilon_{k,n}} \right)
\end{equation}
In this equation, $\epsilon_{k,n}$ signifies the eigenvalue of $H_{\text{BdG}}(\textbf{k})$, and $\beta = \frac{1}{k_B T}$, where $k_B$ is the Boltzmann constant. At a specified magnetic field, the magnitude of the pairing gap $\Delta$ is ascertained by minimizing the difference in free energy densities between $F_s[\Delta]$ and $F_n$, where $F_n$ is defined as $F_s[\Delta = 0]$. The attractive interaction strength, $U_0$, is determined by the superconducting critical temperature, $T_c$, at the zero magnetic field, calculated from first principles.

Figure 5b illustrates the calculated in-plane upper critical fields $B_{c2}$ as functions of temperature for monolayer NbSe$_2$F$_2$ and the monolayer NbSe$_2$ CDW phase. At low temperatures, the $B_{c2}$ values for both NbSe$_2$F$_2$ and the CDW phase of monolayer NbSe$_2$ significantly exceed their respective Pauli limits $B_p \approx 1.84 T_c$ (in tesla for $T_c$ in kelvin). Furthermore, at the same temperature, the $B_{c2}$ for NbSe$_2$F$_2$ surpasses that for the CDW phase of monolayer NbSe$_2$. These results unambiguously indicate that NbSe$_2$F$_2$ exhibits enhanced Ising superconductivity that is more resistant to external in-plane magnetic fields.

\section{Conclusions}
In summary, we demonstrate that surface fluorination substantially enhances Ising superconductivity in monolayer NbSe$_2$ by concurrently elevating the superconducting transition temperature and the in-plane upper critical field. We identify three pivotal roles of surface fluorination in this enhancement. First, fluorine atoms are symmetrically and stably adsorbed on both surfaces of monolayer NbSe$_2$, preserving its mirror symmetry and thereby pinning the orientation of carrier spins in the out-of-plane direction. Second, surface fluorination suppresses the CDW state in monolayer NbSe$_2$ and induces a van Hove singularity around $E_F$, resulting in a pronounced increase of DOS at $E_F$, and, consequently, the EPC. Third, fluorine-related low-frequency phonon modes exhibit a strong coupling with the electronic states at the Fermi level, which further amplifies the total EPC strength. Our findings suggest that fluorinated monolayer NbSe$_2$ with a strengthened Ising superconductivity can serve as a promising platform for the exploration of intriguing quantum phenomena and device paradigms such as topological superconductivity and quantum computing.

\section*{Acknowledgements}

We acknowledge the National Natural Science Foundation of China (Nos. 12274013 and 11874079), and the Independent Research Project of Medical Engineering Laboratory of Chinese PLA General Hospital (No. 2022SYSZZKY10). W. S. is supported by the Science and Technology Commission of Shanghai Municipality (STCSM) (Grant No. 22ZR1441800). Y. X. and W. D. are supported by the Basic Science Center Project of NSFC (grant no. 52388201), the National Natural Science Foundation of China (No. 12334003), the National Science Fund for Distinguished Young Scholars (No. 12025405), the Beijing Advanced Innovation Center for Future Chip (ICFC), and the Beijing Advanced Innovation Center for Materials Genome Engineering.


\bgroup
\fixFloatSize{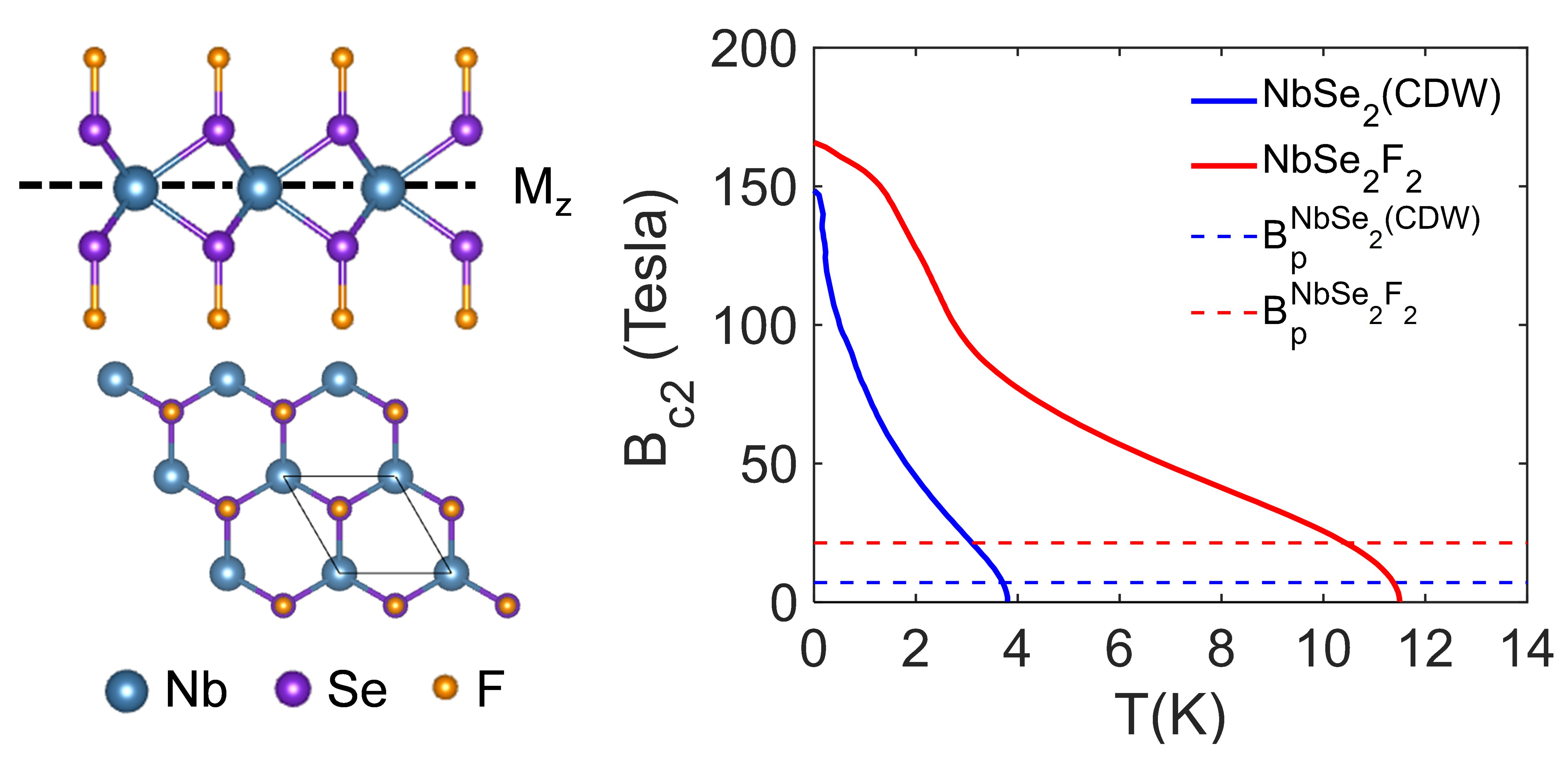}
\begin{figure*}[p]
\centering \makeatletter\IfFileExists{TOC.pdf}{\includegraphics[width=0.8\textwidth]{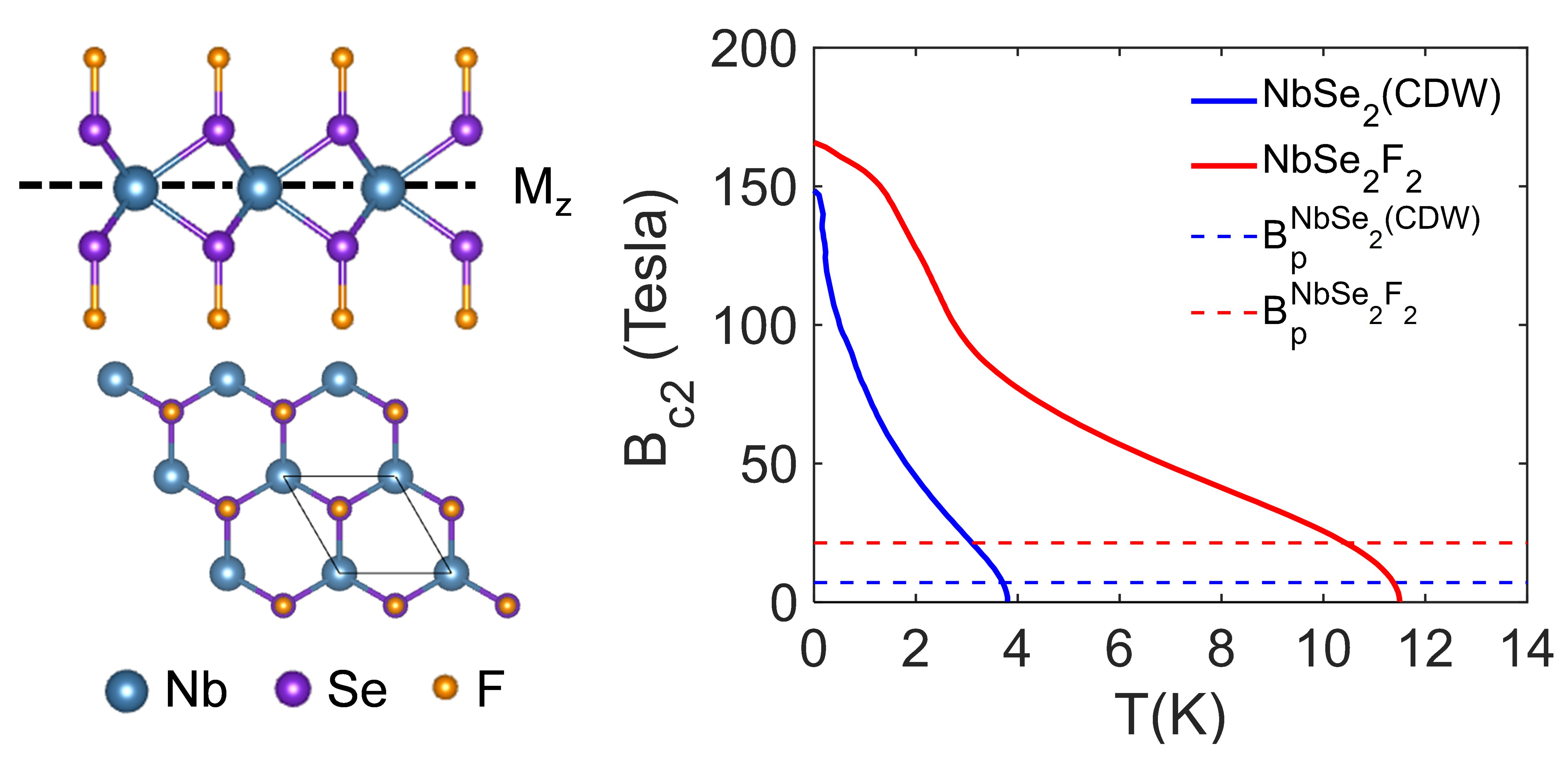}}{}
\caption*{TOC}
\makeatother
\label{TOC}
\end{figure*}

\end{document}